\documentclass[sigconf]{acmart}

\def\BibTeX{{\rm B\kern-.05em{\sc i\kern-.025em b}\kern-.08emT\kern-.1667em\lower.7ex\hbox{E}\kern-.125emX}}
    
\copyrightyear{2019}
\acmYear{2019}
\setcopyright{acmlicensed}
\acmConference[NPSW]{}
\acmDOI{10.1145/1122445.1122456}
\acmISBN{978-1-4503-9999-9/18/06}

\usepackage{url}
\usepackage{graphicx}
\usepackage{amsmath}
\usepackage{algorithm}
\usepackage{algorithmic}
\usepackage{latexsym}
\usepackage{multirow}
\graphicspath{{./figs/}}

\usepackage{pdfpages}
\begin{document}

%
\title{Firewall Regulatory Networks for Autonomous 
Cyber Defense}

%
\author{Qi Duan}
\email{qid@andrew.cmu.edu}
\affiliation{%
  \institution{Carnegie Mellon University }
  \country{U.S.A}
}
\author{Ehab Al-Shaer}
\email{ehab@andrew.cmu.edu}
\affiliation{%
  \institution{Carnegie Mellon University}
  \country{U.S.A}
}

\begin{abstract}
There are multiple intrinsic features that show the lack of resiliency of the current network access control
 architectures, especially firewalls, against dynamic and sophisticated cyber attacks.
 These features include (1) {\em lack of feedback support--} the difficulty to obtain the global
 knowledge of all firewalls in the system as different firewalls are often managed by
 different groups int he same enterprise, (2) {\em nonadaptivity}-- manual firewall rule policy configuration
 is time consuming, error-prone and not effective in real-time attack response,
 (3) {\em passiveness--} the lack of mutual interactions among firewall devices, even though each has a limited visibility based on the subsets of the managed assets, and
 (4) {\em lack of optimal emergent decision-making--} human-in-the-loop management makes the diagnosis of misconfiguration,
 conflict resolution, and timely responding to increasing risk infeasible or highly inaccurate.

  In this paper, we present  the principles of designing new self-organising and autonomous management protocol to govern the dynamics of bio-inspired decentralized firewall architecture based on Biological Regularity Networks. The new architecture called Firewall Regulatory Networks (FRN) exhibits the   following features
  (1) automatic rule policy  configuration with provable utility-risk appetite guarantee,
  (2) resilient response for changing risks or new service requirements, and
  (3) globally optimized access control policy reconciliation.
  We present the FRN protocol and formalize the  constraints  to
  synthesize  the undetermined components
  in the  protocol  to produce interactions that can achieve these objectives. 
  We illustrate  the feasibility  of the FRN architecture
  in multiple case studies.
\end{abstract}

\maketitle



\section{Introduction}
\label{sec:intro}

The research community in the domain of cyber security strives to build resilient and robust architectures for cyber security which can have self-awareness, can reconfigure itself autonomously and can efficiently perform risk mitigation at global level with limited human intervention. To meet these objectives, even after decades of development of cyber security systems, there still exist multiple intrinsic features that make the maintenance of  cyber security devices, especially firewalls, not scalable and allow adversaries to not only plan and launch attacks effectively but also learn and evade detection easily. These features  include (1) {\em lack of feedback support}: the difficulty to obtain the global
 knowledge of all firewalls in the system as different firewalls are often managed by
 different groups int he same enterprise, (2) {\em nonadaptivity}: manual firewall rule policy configuration
 is time consuming, error-prone and not effective in real-time attack response,
 (3) {\em passiveness}: the lack of mutual interactions among firewall devices, even though each has a limited visibility based on the subsets of the managed assets, and
 (4) {\em lack of optimal emergent decision-making}: human-in-the-loop management makes the diagnosis of misconfiguration,
 conflict resolution, and timely responding to increasing risk infeasible or highly inaccurate.

As an example, in a simple network, a web service
is behind firewall 1, a database service is behind firewall 2, and firewall 3 is
connected to Internet. The web service access may involve database access.
The network administrator needs to open the web service to
the Internet. Now  we have four choices under different risk and utility requirements:
\begin{itemize}
  \item {Deny the web access in firewall 1.}
  \item {Allow the web access in firewall 1 and  database access in firewall 2.}
  \item {Allow the web access in firewall 1 and deny  database access in firewall 2.}
  \item {Allow the web access in firewall 1 and partially allow  database access in firewall 2
  (restriction).}
\end{itemize}

This means that one have four access possibilities for the flow related to the
web access. In general if we have $n$ dependent flows and every flow has $N$ possibilities,
the number of  possible combinations will be    $N^n$, which may not be scalable for
a central controller to analyse and manage the firewall rule sets.

Careful inspection of nature enlightens the fact that in human body, every functionality is performed and controlled at cellular level. Cells compose dynamic systems of complex interacting networks in which proteins, genes and small regulatory molecules play together in a programmed manner to perform multiple tasks in an organism. Some proteins have the function of regulating the expression of genes by turning them on or off. This process of interaction, between genes and protein regulatory elements, establishes a Biological Regulatory Network (BRN). Therefore, every function in the body at organismic level is dependent upon BRNs \cite{frn_24}. Not only the functionality, but also the evolution of the morphological (structural/physical) features is highly influenced or controlled by the behavior of BRNs at cellular level. BRNs often contain feedback loops in order to impose a controlled mechanism intended to maintain an optimal concentration of proteins in a cell at global level. Whenever a threat or perturbation arises in the environment, which can lead system to a disease state, the regulating entities in a BRN regulate each other to eliminate or mitigate the threat by maintaining an optimal concentrations of proteins in a cell to meet the global objective with reconciliation. This biological phenomenon can be summarized in a simple way that the dynamics of the living system is controlled by BRNs, and at any given time, BRN of a living organism should optimize the global cell behavior by maintaining the concentrations of proteins, at local level, to make it survive in its (abnormal) environmental conditions. The characteristics of the biological systems which yield such a defense mechanism are as follows: (1) interactive-nature: (helps sensing information from neighboring entities), which includes interactions at cellular level and interactions between genes which belong to the BRN;  (2) self organization, which includes self awareness (sense making by analyzing already sensed information), decision making (evaluation $\&$ realization of risk/threats from external environment), and dynamic nature (changing internal states dynamically and autonomously in accordance with evaluation of threats from external environment and firing/triggering necessary actions to allow/restrict/limit some behavior); and (3) emergent behavior, that is, the emergence of unanticipated complex global behavior from basic set of rules, which includes emergence of immune system, tendency to maintain normal behavior or remain in progressive cycle to avoid deadlock in BRNs, etc.

Although a number of bio-inspired networking techniques have been proposed, they are mostly engineered to provide reliability or resilience against specific threats or attacks, and they do not constitute a scientific ground for creating reasoning frameworks based on understanding the benefits and limitations of bio-inspired networks.
In this paper, we
investigate the rules and dynamics
 governing biological systems and apply them to build bio-inspired Firewall Regulatory Networks (FRNs) with the
  following features (1) automatic rule policy  configuration with provable utility-risk appetite guarantee,
  (2) resilient response for changing risks or new service requirements, and
  (3) globally optimized access control policy reconciliation.

The incentive of every firewall in a cyber infrastructure is to reduce the risk and  increase the usability and demand of the assets, for which it is responsible. In large scale cyber systems multiple firewalls are intertwined, and security policies of any firewall are not designed to reinforce neighboring  firewalls, rather they are more focused towards the specified interests (usability and demand) of the important assets behind them.  Consequently, one wrong action performed by an operator (to fulfill demand) at local level in any firewall, might have catastrophic global impact (by increasing risk on the other devices/assets), which is hard (or impossible) to comprehend via manual configuration.   The important questions which remain to be answered are: (1) is it possible to come up with provably correct set of global policies which can represent the interest of each individual firewall? (2) how to achieve the correct global policies through autonomous interactions without human intervention? (3) can the interactions be done efficiently in real time? (4) how to make it scalable with large-scale networks?

In this paper we address these challenges.
The major  contributions of the paper include (1) establishing the framework of BRN-inspired
cyber system, called Firewall Regulatory Network (FRN), which can achieve self organization, automatic conflict resolution, re-configuration, and resilience against environmental change and adversary behavior, (2) the formal protocol of mutual interactions among devices inside a FRN, and (3) development
of the FRN synthesizer and formalization  of  security
and mission constraints which can lead to the  resolution
of satisfiable interactions.
The FRN synthesizer aims to generate
a correct interaction mechanism  that includes the type and magnitude
of the interactions at a
certain point of time among all  firewalls (to resolve
reconciliation issues), so that bad/malicious states can be
avoided and mission requirements can be satisfied. For this
purpose, we formalize the synthesis
as a Constraint Satisfaction Problem (CSP).

The main tool we use for FRN  formalization  is  Satisfiability Modulo Theories (SMT). SMT is a powerful tool to solve constraint satisfaction problems arise in many diverse areas including software and hardware verification, type
inference, extended static checking, test-case generation, scheduling, planning, graph
problems, etc.~\cite{BM09}. An SMT instance is a formula in first-order logic, where some function and predicate symbols have additional interpretations. SMT is the problem of determining whether such a formula is satisfiable~\cite{DLL62,DP60,Ganesh07}. SMT  provides a much richer modeling language than is possible with  SAT~\cite{DP60,GJ90}. Modern SMT solvers can check formulas with hundreds of thousands variables and millions of constraints~\cite{MB09}. The SMT based formalizations are flexible and  can take additional constraints from real applications conveniently.

\section{Background and Related Works}
\label{sec:bio-related}

A gene regulatory network or genetic regulatory network (GRN) is a collection of regulators interact with each other and with other substances in the cell to govern the gene expression levels of mRNA and proteins. The regulator can be DNA, RNA, protein and their combination. The interaction can be direct or indirect (through their transcribed RNA or translated protein)~\cite{frn_24}.

As an example, suppose there is an input viral infection signal. When the body is infected with a virus the receptor proteins attached to the cells detect the virus and send a stimulus to a class of proteins called ``transcription factors''.
The transcription factors signal the activation of the protein synthesis in order to create proteins that will kill the virus and fight the infection.
At the beginning of the protein synthesis procedure the DNA strand unfolds  and creates an mRNA strand. The mRNA strand is the compliment to the DNA sequence which is basically a list of ingredients needed to build the protein.
So with the mRNA we have the ingredient list, and the body goes to fetch the building blocks that are on the list of ingredients in order to make a sequence of amino acids - a sequence of amino acids is folded to the final structure of the protein.
There are only 20 amino acids in human body, and after the amino acids  are combined together the protein is made and taken to the source of the infection.
Once the infection is removed and cleared an inhibitory signal is sent to the site of protein synthesis to order the DNA to stop coding proteins and refold back again and store it away and at this point the system is in homeostasis.

A preliminary work of our investigation of Bio-inspired networks can be found in~\cite{frn-report}.
However, the framework in this paper is completely different.  
\section{FRN Architecture Framework}
\label{sec:arch}

\subsection{System Components}


In the framework, every FRN contains multiple Bio-inspired firewalls, and every firewall has its own decision engine which contains risk-aware rules/policies, and takes input from receptors, sensors, traffic logs, external events (such as attack alarm), or action signals from other bio-inspired firewalls. The outputs of the decision engine are called actuators, which can be either activators or inhibitors. The output signals are sent to designated peer firewalls. There also exists a feedback mechanism to regulate the interactions among bio-inspired
firewalls.  We believe that FRN can achieve more resiliency and adaptivity than existing cyber infrastructures.  The FRN is distributed without central control, yet can achieve self-management, self-organizing and resiliency. The bio-inspired firewall rules can be dynamic and non-deterministic. The active  firewall rules in a specific time is regulated by the bio-inspired  decision engine, which controls the on/off state of the  rules based on external feedback (from sensors and other bio-inspired firewalls) and system requirements. The
firewall actions include deny, allow, packet inspection, etc.

\subsection{Feedback Control}
\label{sec:feed}

Self-organizing and stable entities such as cells, organs, organisms, societies and sufficiently complex machines must constantly receive information not only about the external environment (via input variable measures by sensors) but also about the state of some of their own elements. The information is then used to make suitable adjustments (via internal functions). This is called feedback or retroaction, operating on internal functions. The basic principle is that any undesired deviation ($\delta_X$) in the value of a variable ($X$) triggers the readjustments of $X$ itself. An element can either affect its own synthesis (evolution) or via series/ chain of interactions with other elements, e.g. $X$ may affect the evolution of $Y$, and $Y$ affects the evolution of $Z$ which in turn affects $X$. Elements connected by such a closed chain for the purpose of influencing/regulating each other, form a feedback loop. The elements of feedback loop do not subject to other interactions, as each element is directly influenced/regulated from its immediate predecessor/ predecessors, and the effect of regulation travels throughout the chain, which helps controlling the optimum values for different variables and becomes the reason for maintaining normal states. A simple n-element feedback loop involves n interactions, each of which can be positive and negative. Positive (conversely: negative) interaction means an element will increase (conversely: decrease) the value of a certain parameter of its follower (one such important parameter can be thought as the value of risk associated with that element).  The major simplification obtained from the  observation is that in any simple feed-back loop an element either exerts positive or negative effect on its own evolution.  Therefore, feedback loops can be classified into two classes: positive and negative. This fundamental fact about the nature of interactions (which is obvious) can prove to be very important to control and maintain the normal behavior of the system/network/cyber system.

The major difference between BRN and FRN is that in FRN we
need to find the level of increase/decrease (in units)
for activation/inhibition since the change of any number of units
of access control can be done immediately without delay, where
in BRN one interaction of activation/inhibition is assumed to be
increase/decrease of one unit.


\subsection{FRN Synthesis Framework and Protocol}
\label{sec:syn}

The FRN  synthesis
engine is built on the FRN protocol and
 uses system topology, service specifications,
mission properties,
and risk/utility metrics   as  input.
 To  achieve a correct feedback control mechanism (for
self organization), one needs to synthesize
the set of interactions at a
certain point of time among all  firewalls (to resolve
reconciliation issues), so that mission properties can be
achieved.
The goal of synthesis is to
solve the constraint satisfaction problem to find the satisfiable
interaction signals (or responses) for every individual firewall,
given possible input (signals from neighboring firewalls). More specifically,
one needs to find the interaction type (activation or inhibition) and number of units for the
interaction.

We use access control vector (ACV) to denote the
status of the firewalls in  FRN.
ACV is an integer vector denotes the status of individual  firewalls.
Here the ACV
in a FRN has the format
\[<\tau^1_{1}, \ldots, \tau^m_{1},| \ldots |\tau^1_{n}, \ldots, \tau^m_{n}>\]
where $n$ is the number of firewalls and $m$ is the number
of rules in every firewall  (for simplicity, we assume that
every firewall has the same number of rules and all rules
in the same firewall are independent), and
$\tau^j_{i}$
denotes the  access control level of
 rule $j$ in firewall $i$ (such as access, deny, inspection, etc).

The service specifications include the
source, destination, utility, risk, and CVSS scores
of every service. The mission properties include the
risk and utility requirements  of the services
and related firewalls of the
mission. The risk of a service is denoted as service (or global)
risk and the risk of a specific firewall
is denoted as firewall risk.
 For a given service $s$, suppose the maximum
 utility can be achieved is $U_{m_s}$, then
 utility of $s$  can be defined as the percentage of reachable flows
 over total possible flow associated with the
 service times $U_{m_s}$, that is
   \begin{align}
     U_{s} = \frac{\text { No. of reachable flows } }
     { \text{ total possible flows }  } \cdot U_{m_s} 
     \end{align}
     The service utility of a mission
     is the weighted summation of all
     utility of all services included in the mission.
     The firewall utility induced by a service
     is defined in the system specification, and
     the total firewall utility  is the
     summation of all utility services related to the firewall.

     Given a host $j$ and service $s$, if we assume that the possible attacks from any other hosts  against
it are all independent, then the
probability that at least one host can  attack $j$ induced by service $s$
is
  \begin{align} 
  1- \displaystyle \prod^n_{i=1} (1-L^s_{ij})  
  \end{align}
 where
  \begin{align}
   L^s_{ij} = w^s_i \cdot (1- \Gamma^s_{ij}) 
   \end{align}
    and $w^s_i$ is the CVSS  score and  $\Gamma_{ij}$ is the
  resistance of service $s$ between host $i$ and $j$. If $i$ cannot reach $j$ (for example,
  blocked by a firewall rule)
  then $\Gamma^s_{ij}$ is 1. If $i$ can reach $j$, then  $\Gamma^s_{ij}$
  is 0.
  So the risk of a host $j$ caused by service $s$ can be estimated as
     \begin{align} 
      R^s_j =( 1- \displaystyle \prod^n_{i=1} (1-L^s_{ij}) )  \cdot I_j
      \end{align}
      
      where $I_j$ is the attack impact of host $j$.
      For a firewall $k$, the local risk induced by service $s$ is the
      summation of the risk of all  hosts that are behind it and related
      to $s$.
       That is,
      \begin{align} 
       R^s_{k} = \displaystyle \sum_{ \text{ j behind k and related to s } } R_j 
       \end{align}

   The total risk of firewall $k$ is the sum of  risks induced by all services.
      \begin{align} 
       R_{f_k} = \displaystyle \sum_{s}  R^s_k 
       \end{align}

      The service  risk associated with $s$ is the summation of
      firewall  risks related to $s$, that is
      \begin{align} 
          R_{S_s} = \displaystyle \sum_{k}  R^s_k 
          \end{align}

We define an activation signal between rule $r$
of two firewalls
$i$ and $j$ as $a^r_{ij}$ and an inhibition signal
between $i$ and $j$  as $h^r_{ij}$.

The trigger for an activation can be opening a service (adding
rules) or another activation/inhibition. The trigger
for an inhibition  can be
(1) user actions that violate security policies,
(2) adding services with high risk or
there is not enough resource for the service,  (3) detection of
malicious behaviors, or (4) another inhibition or activation
from its neighbors.

Next we consider the details of the four possible types of induced interactions:
\begin{itemize}
\item{An activation causes another activation: $a^r_{ij} \Rightarrow a^{r'}_{jk}$.}
\item{An activation causes another inhibition: $a^r_{ij} \Rightarrow h^{r'}_{jk}$.}
\item{An inhibition causes another inhibition: $h^r_{ij} \Rightarrow h^{r'}_{jk}$.}
\item{An inhibition causes another activation: $h^r_{ij} \Rightarrow h^{r'}_{jk}$.}
 \end{itemize}

Note that the rule $r'$ in the induced interaction can be the same
or different as  rule $r$.
All the four types of induced interactions can happen in
reality. The first type of induced interactions can happen when
a new service is opened, the firewalls in the path from the
source and destination will be notified  one by one (propagated)
to activate the access. For example, accessing the web server leads to enable access to the Kerberos authentication sever.
The second type will happen when a new
   service is opened and notified for a firewall, the firewall
   detects that it will introduce high risk then it will send
   related inhibition signal.
   The third type will happen
   when the inhibition signal needs to be propagated.
   For example,
   if a user is revoked access to the server, then his/her machine will be automatically inhibited from sending traffic to the server.
 The last type will happen when a
   high risk service flow is denied, a firewall may activate
   another service flow to achieve desired utility.
   This means that risk can still be decreased
   with the utility  increased or maintained.
   For example, inhibited  access
   to the server via one path due to DDoS leads to activate another path to access the same server.

  Suppose a new service is opened then the
protocol to propagate the access of the service from source to destination
can be described as follows
\begin{align}
\nonumber
\forall i,j, \,(r.src \in i)\land path(j,r.dest) 
\land connect(i,j) \\
 \land (util(r.service) \geq T_{U_i}) \Rightarrow a^r_{ij}  
\end{align}
\begin{align}
\nonumber
 \forall k, \, a^r_{ij} \land path(r.src,k) \land path(k,r.dest) \\
\land connect(j,k)  \Rightarrow a^r_{jk}  
\end{align}
\begin{align} 
 \forall k, \, a^r_{ij} \land (r.dest \in j) \Rightarrow \neg  a^r_{jk} 
\end{align}

  where $i,j,k$ are firewalls, $r.src$, $r.dest$ and $r.service$ are the source,
destination and service of rule $r$ respectively,
$util(r.service)$ is the utility of the service, $T_{U_i}$
is the utility threshold,  and connect(i,j) denotes whether
firewall $i$ is directly connected to $j$. The first equation  is the
activation signal from the source to the next hop, the second
equation specifies all subsequent activations. The last
  equation guarantee that no further activation will be
  signaled when the destination is reached.

  If a firewall detects that the risk of allowing a service is
  higher than threshold, then it will send an inhibition signal. That is,
 \begin{align}
 \forall i,j, a^r_{ij} \land (risk(r.service) \geq T_{R_j}) \Rightarrow h^{r'}_{ji}  
 \end{align}
where $T_{R_j}$ is the specified threshold.

  Note that it is important to select the appropriate $r'$ to make sure the
  risk will below threshold and at the same time minimize utility loss.
  We call the problem as Optimal Policy Risk Mitigation (OPRM).
  This is a knapsack problem  (or fractional
  knapsack problem if a flow can be partially restricted) which exists efficient
   approximation algorithms.

  An inhibition signal may cause the dependent inhibition signals, that is,
  \begin{align} 
  \forall i,j,k, h^r_{ij} \land depend(r,r') \land (r' \in k) \Rightarrow h^{r'}_{jk}   
  \end{align}
  where $depend(r,r')$ means that $r$ and $r'$ are dependent.

  An inhibition signal may cause activation signal, that is,
  \begin{align}
   \forall i,j,k,  h^r_{ij} \land (risk(r'.service) \leq T_{R_j}) \land  (r' \in k)   \Rightarrow a^{r'}_{jk}  
   \end{align}

   It is also important to select the appropriate $r'$ to make sure the
  risk  is lower than specified threshold and at the same time maximize utility gain.
  We call the problem as Optimal Policy Utility Restoration (OPUR).
  This is also a knapsack problem  or fractional
  knapsack problem.

 To resolve conflict in the interactions, we set the
   priority for conflict resolution, that is
   global risk $>$ local risk $>$ global utility $>$ local utility.
   We will show how this works in the case studies.


\subsection{FRN Synthesis Formalization}

The FRN synthesis
 are done
  in two steps. In the first
step the synthesizer  needs to find the satisfiable
ACV corresponds to the utility/risk change. In the second step, the controller needs
to find the satisfiable interactions (among the firewalls)
that change the old ACV to the new ACV.

When adding a new service,
 the satisfiable ACV configuration can be solved with the following constraints:

 (1) The added service should be reachable from its source to destination. Reachability
 can be expressed as a DNF (Disjunctive normal form)
 of all possible paths, if we assume that the ACV only contains
 binary values (allow or deny). That is
  \begin{align}
   \bigvee \bigwedge \tau^r_i 
   \end{align}
  where   $\tau^r_i$ is the ACV value of rule $r$ (related to the service)
  in firewall $i$, and the logical $OR$ operation is for all possible paths
  for the service, the logical $AND$ operation is for the firewalls in the
  path.

 (2) The  risk of any firewall $i$  after adding the service
  should be less than the specified threshold.
  
   (3) The service risk  caused by adding the service $s$
  should be less than the specified threshold.

  (4) The utility of the firewall $i$ by adding the service should  be greater than
  specified threshold.

  (5) The  service utility by  adding the service $s$ should greater than
  specified threshold.

Suppose there are $n$ firewalls  $f_1,\ldots, f_n$, and
every service (or flow, for simplicity, we
assume every service only has one flow)  has a fixed source
and destination,
and for every firewall  $f_k$, there are
$m_k$ possible input (interaction) signals  $S^k_1,\ldots, S^k_{m_k}$, and
$\eta_k$ possible interaction responses $A^k_1,\ldots, A^k_{\eta_k}$.
Here the number $\eta_k$ is determined by the number of
rules and the number of neighbors of the firewall. If the
firewall has $n_1$ rules and $n_2$ neighbors, then the firewall
may have at most $2n_1n_2$ (coefficient 2 means there are two possibilities,
activation or inhibition) types of interactions.
If we use Boolean
$b^v_{uk}$  to denote at step $v$ whether $f_u$
takes interaction response $A^u_k$ or not, then
we have the following constraints.


 \begin{align}
  \forall u,v,k, \,\, b^v_{uk} \in \{0,1\} 
 \end{align}

The response $A^u_k$ will cause the associated
ACV change according to the FRN protocol, that is
   \begin{align}
    \forall u,v,k, \,\, (b^v_{uk} = 1) \Rightarrow \tau^{rv}_{u'} = \tau^{r(v-1)}_{u'} + \delta^{rv}_{u'} 
    \end{align}
   where $\tau^{rv}_{u'}$ is the
   access control value of rule $r$ associated with interaction  in
   $f_{u'}$ at time step $v$,  $u'$ is the index of the  interaction peer of firewall
    $f_u$ and   $\delta^{rv}_{u'}$ is the change of
   access level of rule $r$ caused by interaction $A^u_k$ at step $v$.
   If the signal is activation, $\delta^{rv}_{u'}$ is  positive
   otherwise it is negative. That is
   \begin{align}
    A^u_k \in ACT_{u} \Rightarrow (\delta^{rv}_{u'} >0 )  
    \end{align}
  \begin{align}
   A^u_k \in INH_{u} \Rightarrow (\delta^{rv}_{u'} <0 )  
   \end{align}

   where $ACT_{u}$ and $INH_{u}$ are the set of possible received activation
   and inhibition signals
   of firewall $f_{u}$, respectively.

Also, we require that after a limited number of steps
the result ACV is the desired one.
That is
\begin{align}
\forall r,k,  \,\, \tau^{r\Delta}_k = \tau^r_k  
\end{align}
  where $\Delta$ is the threshold number
  of steps for the convergence and $\tau^r_k$
  is the desired ACV value of rule $r$ in firewall $f_k$
  solved from the first step of synthesis.


Additionally, we can include constraints
for the protocol. First, we can assume that
only neighbored firewalls can interact each other,
that is
    \begin{align}
     \forall i,j,v,r, \neg connect(i,j) \Rightarrow a^{rv}_{ij} = h^{rv}_{ij} =
    a^{rv}_{ji} = h^{rv}_{ji} = 0
    \end{align}
here $a^{rv}_{ij}$ and $h^{rv}_{ij}$ are the
same variable to denote whether activation or inhibition
happens between $f_i$ and $f_j$ for rule $r$ as the description
in the protocol, with additional notation $v$ to denote
time step.

Also, a firewall will not send the same activation (or inhibition) to the same firewall
again after receiving a response.
That is,
  \begin{align} 
  \nonumber
  \forall v,r,i,j,  (a^{rv}_{ij} =1) \land
    (a^{r(v+1)}_{ji} =1 \lor h^{r(v+1)}_{ji} =1)  \\
    \Rightarrow
     \displaystyle \, \forall_{v'\geq v+2} (a^{rv'}_{ij} = 0) 
      \end{align}
   \begin{align} 
   \nonumber
   \forall v,r,i,j,  (h^{rv}_{ij} =1) \land
    (a^{r(v+1)}_{ji} =1 \lor h^{r(v+1)}_{ji} =1) \\
     \Rightarrow \displaystyle \forall_{v'\geq v+2} \, (h^{rv'}_{ij} = 0) 
      \end{align}

Note that all
the synthesis is done off-line but the interactions
will happen in real time response.
The limitation of the synthesis is that one cannot
find a solution for all possible utility/risk changes,
however, we can solve the  interactions
that converge to the satisfiable configurations for
those important or high priority utility/risk changes.
Also note that the synthesizer is different
from a central controller since the synthesizer
only tries to find the correct interactions off-line
for a given set of utility/risk changes, but not do the
ACV configuration change directly. In a system with central controller
all configuration changes can be done by the controller
directly. However, this is not practical
for  large systems that need real time responses to
many types of utility/risk changes. The FRN can
achieve the global required coherence through local interactions,
and the main challenge is to find the correct interactions
with desired provable properties. We can consider every
firewall as a finite state machine which generates
certain output interaction signals given specific input,
and the task of the synthesizer is to find these correct
 transitions for the finite state machines with provably
 correct properties.

\section{Conclusion and Future Work}
In this paper we  present a new paradigm
of cyber security in which that a network of
firewalls called FRN that can achieve global coherence through local
interactions. We design FRN protocol to achieve the
mission objectives of risk and utility adaptively and
  formalize the constraints to solve the satisfiable interactions among
  firewalls. We implement the FRN protocol and formalization and illustrate the
  feasibility  the FRN architecture through 
  multiple case studies.

Many interesting future directions can be extended from the work
in this paper. The first direction is to consider the model without
off-line global synthesizer and every firewall
 needs  to do its own synthesis.
  In this case,  one must guarantee that
  all  global  constraints  can  be
decomposed into local ones so that the synthesis
are  done  on  each  firewall
separately, and the combination of local constraints
satisfy the global  constraints.
For every individual firewall, the inputs to the synthesis
program include system configuration, current sensing
information, and local constraints. The verification
procedure can verify that all global constraints
are satisfied from the local constraints.
We plan to find solutions to leverage firewall interactions
and feedback loop control based on the
firewall interdependencies through
system decomposition.
   If the system has $N$ firewalls, each can take any
of the $M$ local states, then there may be as many as
$M^N$ states in
the whole system.
The search space to find a certain satisfiable
 configuration is thus exponential. On the other hand, if we
 decompose the system based on every local firewall
  and its local impacting/impacted
 bio-inspired entities, we divide the system
into $N$ subsystems each having around $d$
interdependent firewalls, and $d$ is usually
small. In this case every subsystem
has $(d+1)^M$ possible states, which
is a much smaller.

Another future direction is  to apply game theory
methods in FRN to achieve the global requirements through
local interactions which can be
modelled as a cooperative  game  where the equilibrium is reached at a point with maximum global benefit (in terms of achieving the desired resiliency or the effectiveness of defending attacks), or a Stackelberg game that one or more firewalls take some initial actions to trigger subsequent actions from other firewalls to achieve the best benefit.
However there are two
major difficulties here. First, the game may not be scalable, which
means it will be infeasible to find the desired equilibrium. Second,
modelling the whole system as a game is not flexible
for dynamic system changes  and
it is difficult to incorporate additional global constraints
into the game.
It is interesting to find solutions to
overcome these difficulties.
Methodologies
 used to solve differential games may also be
 used in FRN  synthesis. Differential games solve a group
  of problems related to the modeling and analysis of
  conflict in the context of a dynamical system.
  More specifically, a state variable or variables evolve
  over time according to a differential equation.
  However, differential games are conceptually far more
complex than optimal control problems in the sense that it is no longer
obvious what constitutes a satisfiable solution in a FRN
since FRN synthesis may  involve
  conflicting local objectives, so one needs further investigation to apply
  methodologies in  differential games.

\bibliographystyle{plain}
\bibliography{main}
\end{document}